\documentstyle[floats,twocolumn,aps,pre,amsfonts,amssymb]{revtex}

\input{epsf}

\newcommand{\rsn}{R_{\text{SN}}}              
\newcommand{\mean}[1]{\langle #1\rangle}      
\newcommand{\Omres}{\Omega_{\text{res}}}      
\newcommand{\Tres}{T_{\text{res}}}            
\newcommand{\hS}{\hat{S}}                     
\newcommand{\Omax}{\hat{\Omega}}              
\newcommand{\sigmax}{\sigma_{\text{max}}}     
\newcommand{\phasd}[1]{\chi^{(#1)}}           

\begin{document}
\draft

\title{Stochastic Resonance in Neuron Models: 
       Endogenous Stimulation Revisited}
\author{Hans E.\ Plesser$^{1,2,}$%
        \thanks{Corresponding author: hans.plesser@itf.nlh.no}
        and Theo Geisel$^{1}$}       
\address{$^1$Max-Planck-Institut f\"ur Str\"omungsforschung
         and Fakult\"at f\"ur Physik, Universit\"at G\"ottingen\\
         Bunsenstra\ss{}e~10, 37073~G\"ottingen, Germany}
\address{$^2$Institutt for
        tekniske fag, Norges landbruksh{\o}gskole, P.O.Box 5065,
        {\AA}s, Norway}
\date{To appear in Phys Rev E}
\maketitle


\begin{abstract}
  The paradigm of stochastic resonance (SR)---the idea that signal
  detection and transmission may benefit from noise---has met with
  great interest in both physics and the neurosciences.  We
  investigate here the consequences of reducing the dynamics of a
  periodically driven neuron to a renewal process (stimulation with
  reset or endogenous stimulation). This greatly simplifies the
  mathematical analysis, but we show that stochastic resonance as
  reported earlier occurs in this model only as a consequence of the
  reduced dynamics.
\end{abstract}




\section{Introduction}\label{sec:intro}

The improvement of signal transmission and detection through noise has
been studied keenly over the past two decades under the paradigm of
stochastic resonance; for recent reviews see
Refs.~\cite{Gamm:1998(223),Wies:1998(539)}.  This holds true for the
neurosciences in particular, which have for a long time been puzzled
by the seemingly irregular activity of the nervous system.  A long
series of experiments has now firmly established that sensory neurons
of various modalities benefit from ambient noise
\cite{Long:1991(656),Doug:1993(337),Levi:1996(165),Coll:1996(642),%
Cord:1996(769),Jara:1998(384),Russ:1999(291)}.

Theoretical efforts to explain stochastic resonance in neuronal
systems had to abstract rigorously from the biological complexity of
real neurons to facilitate mathematical
treatment \cite{Zhou:1990(3161),Buls:1991(531),Ging:1995(191),%
Wies:1994(2125),Coll:1996(5575)}.  The
leaky integrate-and-fire model neuron (LIF) \cite{Tuck:Stoc} is likely the
most widely studied of these abstract neuron
models, especially in investigations of
the neuronal code \cite{Gers:1996(76),Mars:1997(735),Troy:1997(971),%
  Bugm:1997(985),Feng:1997(4505),Abbo:1997(220)}.  The main advantages
of this model are its simplicity and lack of memory: Each time the
neuron has been excited sufficiently to fire a spike (output pulse),
it is reset to a predefined state, erasing all memory of past input.
One may thus analyze the intervals between spikes separately, and
build the complete spike train fired by the neuron by concatenation of
intervals.

Time-dependent stimulation complicates this procedure, though, as a
different stimulus is now presented during each interspike interval.
This precludes a straightforward analysis of global properties of the
spike train, such as its power spectral density.  Since stochastic
resonance is typically defined in terms of the signal-to-noise ratio
in response to periodic stimulation, either of the following is
required: (i)~further simplification of the model, or (ii)~the
development of better techniques.

The latter is certainly preferable and has been achieved in the
meantime \cite{Shim:1999(3461),Shim:1999(R33),Ples:1999(7008)}.
Stochastic resonance with respect to both noise amplitude and signal
frequency has been found, and may be relevant to signal processing in
neuronal networks \cite{Ples:CNS00,Ples:PhD}.  Results obtained for
the leaky integrate-and-fire neuron are in qualitative agreement with
those obtained from more realistic, non-linear models \cite{Liu:1999(3453),%
Kana:1999(23)}, indicating that the abstraction to the LIF model was
acceptable.

Albeit stochastic resonance in the leaky integrate-and-fire model may
thus be considered a solved problem, we would like to return here to
some earlier studies \cite{Buls:1996(3958),Ples:1997(228),%
  Shim:1999(3427),Shim:1999(R33)}, which, at least in parts, followed
the simplifying approach~(i).  These studies assumed that the same
stimulus was presented during each interspike interval, i.e., that the
originally periodic stimulus was reset to a fixed phase $\phi_0$ after
each spike.  All intervals are thus equivalent, and global properties
of the spike train may be computed from the interspike-interval
distribution for a single interval using the methods of renewal theory
\cite{Fran:1995(1074)}.  L{\'a}nsk{\'y} coined the terms
\emph{endogenous} and \emph{exogenous} stimulation for stimulation
with and without reset, respectively \cite{Lans:1997(2040)}.

This stimulus-reset assumption is patently unbiological for virtually
all neurons, since it would require the neuron to fully control the
input it receives.  The reset assumption was therefore justified as an
approximation to the full, exogenous dynamics of a periodically driven
noisy neuron along the following lines \cite{Buls:1994(4989)}: If the
neuron is driven by a periodic subthreshold stimulus (a stimulus too
weak to evoke spikes in the absence of noise), then the neuron will
fire all spikes at approximately the same stimulus phase $\phi^*$
after transients have died out.  This phase corresponds roughly to the
phase at which the membrane potential is maximal in the absence of
noise.  Such firing patterns are found in sensory neurons, e.g., in
the auditory nerve \cite{Rose:1967(769)} or cold-receptor neurons
\cite{Brau:1984(26),Long:1996(215)}.  Thus, roughly the same stimulus,
starting from phase $\phi(\tau=0)\approx\phi^*$, is presented during
each interval.  Resetting the stimulus phase to a fixed phase
$\phi(\tau=0)=\phi_0$ will therefore introduce only minor errors.

This reasoning suffers from an essential shortcoming, namely the
choice of a \emph{fixed} reset phase $\phi_0$ independent of both
stimulus and noise.  This reset phase is a free parameter of the
model, which has no counterpart in biological neurons.  We show here
that stochastic resonance in terms of the signal-to-noise ratio as
studied previously in the LIF model with stimulus reset
\cite{Ples:1997(228),Shim:1999(3427),Shim:1999(R33)} occurs because
the model neuron adapts best to the free parameter ``reset phase'' for
a particular noise amplitude.  The signal-to-noise ratio \emph{as
  defined in those studies} diverges monotonically for vanishing noise
if the reset phase is adapted to stimulus and noise in a plausible
way.

The leaky integrate-and-fire model and the methods applied in analysis
are briefly reviewed in Sec.~\ref{sec:model}; see Ref.~\cite{Ples:PhD}
for details. Results are presented in Sec.~\ref{sec:results} and
summarized in Sec.~\ref{sec:disc}.


\section{Model and Methods}\label{sec:model}

\subsection{Leaky integrate-and-fire neuron}\label{sec:if}

The leaky integrate-and-fire neuron model sketches the neuron as a
capacitor with leak current, which is charged by an input current
$I(t)$ until the potential $v(t)$ across the capacitor (membrane
potential) reaches a threshold $\Theta$.  At that instant, an output
spike is recorded and the potential reset to $v_r<\Theta$.  The
assumption that the stimulus is reset after each spike as well
implies that the membrane potential evolves according to
\begin{equation}
  \dot{v}(\tau) = -v(\tau) + \mu + q \cos(\Omega\tau+\phi_0) +\sigma\xi(\tau),
  \label{eq:v}
\end{equation}
in between any two spikes \cite{Tuck:Stoc}; $\tau$ is intra-interval
time, i.e., $\tau$ runs from zero in every interval.  $\mu$ is the DC
component of the stimulus, $q$ its AC amplitude, $\Omega$ the nominal
frequency, and $\phi_0$ the fixed but arbitrary reset phase.  All
quantities are measured in their natural units, i.e., the membrane
time constant $\tau_m$ and threshold $\Theta$.  $v_r=0$ is assumed
throughout.  The noise term in Eq.~(\ref{eq:v}) subsumes both
biochemical and network noise \cite{Manw:1999(1797),Main:1995(1503)},
and is taken to be Gaussian white noise
[$\mean{\xi(t)\xi(t')}=\delta(t-t')$].  A different realization of the
noise is used for each interval.  Sub-threshold stimuli are
characterized by $\sup_{t\to\infty} v(t) = \mu+q/\sqrt{1+\Omega^2}<1$
for $\sigma=0$.  We restrict ourselves here to these stimuli, because
they appear to be more relevant for the encoding of periodic signals
in sensory systems than superthreshold stimuli
\cite{Gers:1996(76),Kemp:1998(1987)}.  

\begin{figure}
  \centerline{\epsfbox{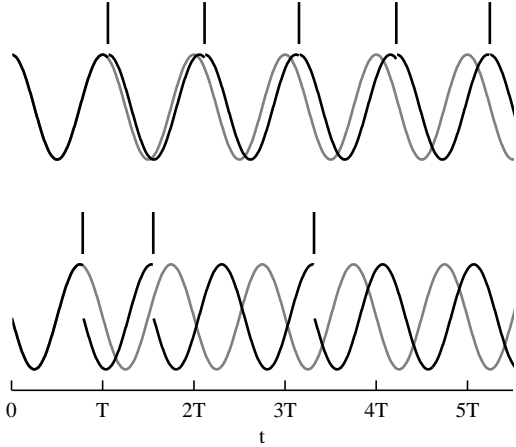}}
  \caption{Effective stimulus and corresponding spike trains for fixed
    reset phases $\phi_0=0$ (top) and $\phi_0=\pi/2$ (bottom) are
    shown in black, while the nominal stimuli are shown in grey.  The
    reset has small consequences in the first case, while the effective
    stimulus differs markedly from the nominal sinusoid for the
    latter.  $T=2\pi/\Omega$ is the period of the nominal stimulus;
    amplitudes are in arbitrary units.  Remaining stimulus
    parameters: $\mu=0.9$, $q=0.1$, $\Omega=0.1\pi$, $\sigma=0.008$.}
  \label{fig:stim}
\end{figure}

The sequence $\tau_1, \tau_2, \dots, \tau_k, \dots$ of intervals
corresponds to an output spike train $f(t)=\sum_k\delta(t-t_k)$ with
spike times $t_k = \sum_{j\le k} \tau_j$.  This spike train is evoked
by an \emph{effective stimulus} consisting of piecewise sinusoids of
length $\tau_k$, as shown in Fig.~\ref{fig:stim}. In contrast, we call
the pure sinusoid $\cos(\Omega t+\phi_0)$ the \emph{nominal stimulus}.
Figure~\ref{fig:stim} indicates that the effective stimulus
approximates the nominal stimulus for a reset phase of $\phi_0\approx
0$, while it is an irregular sequence of piecewise sinusoids for other
choices of the reset phase.  We therefore focus here on $\phi_0=0$ in
accordance with earlier work
\cite{Ples:1997(228),Shim:1999(3427),Shim:1999(R33)}.

\begin{figure}
  \centerline{\epsfbox{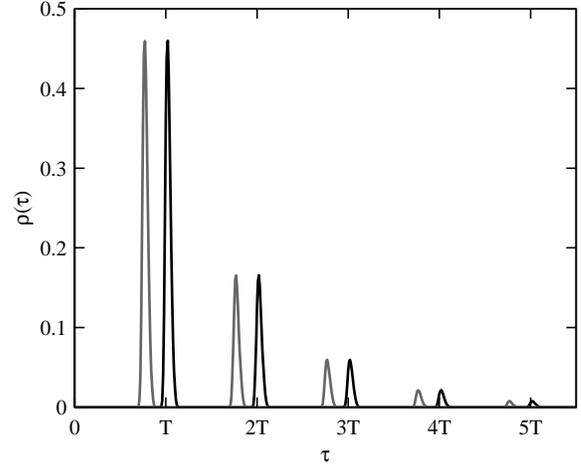}}
  \caption{Interspike-interval distribution in response to periodic
    stimulation with reset to phases $\phi_0=0$ (black) and
    $\phi_0=\pi/2$ (grey).  All else as in Fig.~\ref{fig:stim}.}
  \label{fig:isi}
\end{figure}

Because of the stimulus reset and the whiteness of the noise, all
interspike interval lengths $\tau_k$ are statistically independent,
identically distributed random variables with density $\rho(\tau)$.
The latter can be computed numerically or approximated in closed form
\cite{Ples:PhD,Buon:1987(784),Ples:2000(367)}.  The sequence of
intervals thus forms a renewal process, which is fully characterized
by the ISI density $\rho(\tau)$ \cite{Cox:Theo}.  Periodic
sub-threshold stimulation evokes multimodal ISI densities as shown in
Fig.~\ref{fig:isi} for noise not too strong.  The location of the
first peak depends on the reset phase $\phi_0$, while subsequent peaks
follow at intervals of the nominal stimulus period $T=2\pi/\Omega$.
Comparable ISI distributions are found in sensory neurons
\cite{Rose:1967(769),Brau:1984(26),Lavi:1971(467)}.

\subsection{Signal-to-noise ratio}\label{sec:snr}

The performance of a signal processor is commonly measured in terms of
the signal-to-noise ratio in studies on stochastic resonance.  Since
the spike train elicited from the neuron is a renewal process by
virtue of the stimulus reset, its power spectral density (PSD) is
given by \cite{Fran:1995(1074),Ples:1997(228)}
\begin{equation}
  S(\omega) = \frac{1}{\pi\mean{\tau}}
              \left(1 + 
                    2\Re \frac{\tilde{\rho}(\omega)}{1-\tilde{\rho}(\omega)}
              \right),\;\omega>0,
  \label{eq:S}
\end{equation}
where
\begin{equation}
  \tilde{\rho}(\omega)=\int_0^{\infty} \rho(\tau) e^{i\omega\tau} d\tau
  \label{eq:rhof}
\end{equation}
is the Fourier transform of the ISI density and $\mean{\tau}$ the mean
interspike interval length; note that $\rho(\tau)=0$ for $\tau<0$ by
definition.

\begin{figure}
  \centerline{\epsfbox{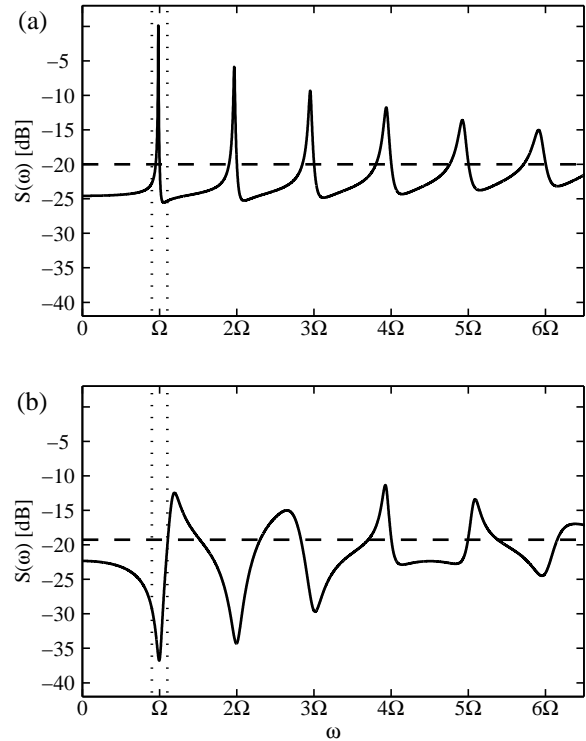}}
  \caption{%
    Power spectral density for
    (a)~reset phase $\phi_0=0$, and (b)~$\phi_0=\pi/2$.  The dashed
    horizontal line is the PSD $S_P$ of a Poisson train of equal
    intensity; vertical dotted lines mark the interval $[0.9\Omega,
    1.1\Omega]$, cf.\ Eq.~(\ref{eq:Smax}).  Note the lack of power
    in the vicinity of the nominal stimulus frequency $\Omega$ for
    reset phase $\phi_0=\pi/2$.  All else as in Fig.~\ref{fig:stim}.}
  \label{fig:spec}
\end{figure}

The input to the neuron is not purely sinusoidal due to the stimulus
reset, and the maximum of the PSD will thus be shifted away from the
stimulus frequency $\Omega$, see Fig.~\ref{fig:spec}(a).  We thus
define the signal as the maximum of the PSD in a window around
$\Omega$ \cite{Ples:1997(228),Shim:1999(3427)}
\begin{equation}
  \hS = S(\Omax) = \max\{S(\omega) | 0.9\Omega < \omega < 1.1\Omega\}
 \label{eq:Smax}
\end{equation}
and refer to the location $\Omax$ of the maximum as \emph{peak
  frequency}.  The signal $\hS$ is undefined if $S(\omega)$ has no
absolute maximum within the window as, e.g., in
Fig.~\ref{fig:spec}(b).  The white power spectrum of a Poissonian
spike train of equal intensity, $S_P= (\pi\mean{\tau})^{-1}$, is used
as reference noise level \cite{Stem:1996(687)}, whence the
signal-to-noise ratio is
\begin{equation}
   \rsn =   {\hS} / {S_p} = \pi \mean{\tau} \hS.
 \label{eq:snr}
\end{equation}

Note that power spectral density and signal-to-noise ratio as defined
above are calculated for infinitely long spike trains.  Any strictly
periodic component of the spike train will thus give rise to
singularities in the PSD.  In the reset model, coherence is broken by
the stimulus reset, resulting in a continuous spectrum
\cite{Prie:Spec,Ples:PhD}.  Due to the different definitions,
signal-to-noise ratios obtained from the LIF model with and without
stimulus reset cannot be compared quantitatively
\cite{Shim:1999(R33)}.  

The spectrum defined by Eq.~(\ref{eq:S}) may have very narrow peaks
for low noise, whence numerical evaluation of
Eqs.~(\ref{eq:S})--(\ref{eq:snr}) may require very high frequency
resolution.

\subsection{Preferred firing phase}\label{ssec:pref}

If the stimulus is not reset after each spike, the probability
$\phasd{k+1}(\psi)$ for the $k+1$-st spike to occur at stimulus phase
$\psi$ can be expressed in terms of the corresponding distribution for
the $k$-th spike as
\begin{equation}
  \phasd{k+1}(\psi) = 
    \int_{-\pi}^{\pi} {\mathcal T}(\psi|\phi)\phasd{k}(\phi) d\phi
  \label{eq:chit}
\end{equation}
where ${\mathcal T}(\psi|\phi)$ is a stochastic kernel
\cite{Tate:1995(917),Ples:1999(7008)}.  For $k\to\infty$, the firing
phase distribution will approach a stationary distribution $\phasd{s}$
and we choose the preferred firing phase as the phase at which the
neuron most likely fires,
\begin{equation}
  \phi^* = \arg\max_{\psi} \phasd{s}(\psi) .
  \label{eq:pref}
\end{equation}

Using this preferred phase as reset phase will yield a viable
approximation to stimulation without reset only if the firing phase
distribution is sharply concentrated around $\phi^*$, and has only a
single maximum.  To ensure this, we require that the vector strength
\cite{Gold:1969(613)} 
of the distribution fulfills
\begin{equation}
  r = \left|\langle e^{i\phi} \rangle\right| = 
      \left|
          \int_{-\pi}^{\pi} e^{i\phi} \phasd{s}(\phi) d\phi
      \right|
      \ge 0.9 .
  \label{eq:r}
\end{equation}
This condition is generally met by the sub-threshold stimuli combined
with weak noise studied here.  Multimodal firing phase distributions
are observed only for slow stimuli in the presence of intermediate to
large noise, and for superthreshold stimuli.  We are therefore not
concerned with complications that may arise from mode-locking as
observed in the latter cases
\cite{Coom:1999(2086),Hunt:1998(1427),Tate:1998(675),Hopp:1982(339)}.


\section{Results}\label{sec:results}

\subsection{Fixed Reset Phase}\label{ssec:fix}

\begin{figure}
  \centerline{\epsfbox{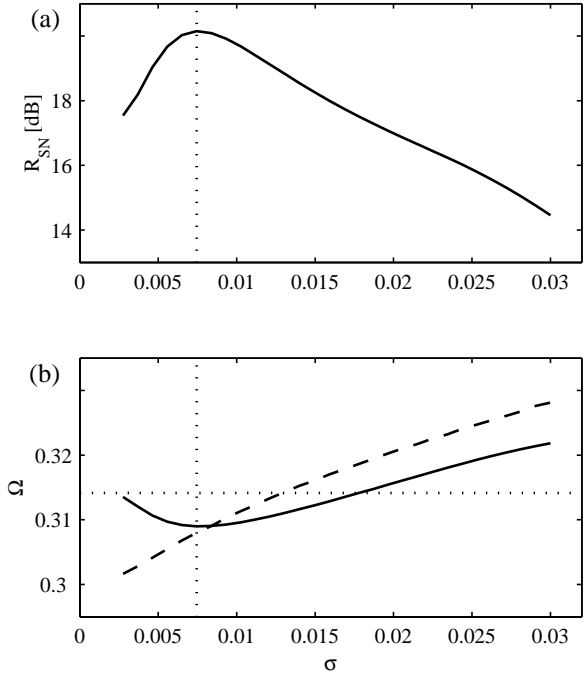}}
  \caption{%
    (a)~Signal-to-noise ratio vs.\ input noise amplitude for fixed
    reset phase $\phi_0=0$.  
    (b)~Peak frequency $\Omax$ (solid) and reset frequency $\Omres$
    (dashed) vs.\ noise amplitude for fixed reset phase.  The dotted
    horizontal line marks the nominal stimulus frequency $\Omega=0.1\pi$,
    while the dotted vertical line marks the optimal noise amplitude.
    Stimulus parameters as in Fig.~\ref{fig:stim}.}
  \label{fig:sr_fix}
\end{figure}

For fixed reset phase, $\phi_0=0$, the model neuron shows typical
stochastic resonance behavior, i.e., a maximum of the SNR at an
intermediate albeit small noise amplitude $\sigmax$ as shown in
Fig.~\ref{fig:sr_fix}(a) \cite{Ples:1997(228)}. The mechanism inducing
stochastic resonance is indicated in Fig.~\ref{fig:sr_fix}(b): the
maximal SNR is reached when the peak frequency $\Omax$ and the reset
frequency $\Omres=2\pi/\Tres$ coincide, where $\Tres$ is the mode of
the ISI density, i.e., the most probable interval between two stimulus
resets.  Coincidence of reset and peak frequencies thus indicates
synchronization between the stimulus reset and the correlations
dominating the power spectrum of the output spike train, yielding
optimal encoding of the nominal stimulus.

This effect may intuitively be explained as follows.  Assume that the
peaks of the ISI distribution are located around $\tau=nT+\epsilon$,
$\epsilon>0$, for a given choice of reset phase and small noise. The
effective stimulus will then be a close to the nominal stimulus, cf.\ 
Fig.~\ref{fig:stim}(a).  The neuron will fire at shorter intervals as
the noise is increased, and for a particular noise amplitude will the
peaks of the ISI distribution be centered about multiples of the
stimulus period $T$.  The effective stimulus will then be reset in
intervals of $nT$.  The neuron is thus on average driven by a periodic
stimulus with period $\Omres=\Omega$, evoking as periodic a spike
train as possible, i.e., one maximizing the signal-to-noise ratio.
This explanation ignores all jitter in spike timing, which reduces the
SNR.  The SNR maximum is thus not found for that input noise amplitude
which yields $\Omres=\Omega$ ($\sigma\approx 0.018$), but for somewhat
weaker input noise, corresponding to smaller output jitter, cf.
Fig.~\ref{fig:sr_fix}(b).

\subsection{Noise-Adapted Reset Phase}\label{ssec:adap}

\begin{figure}
  \centerline{\epsfbox{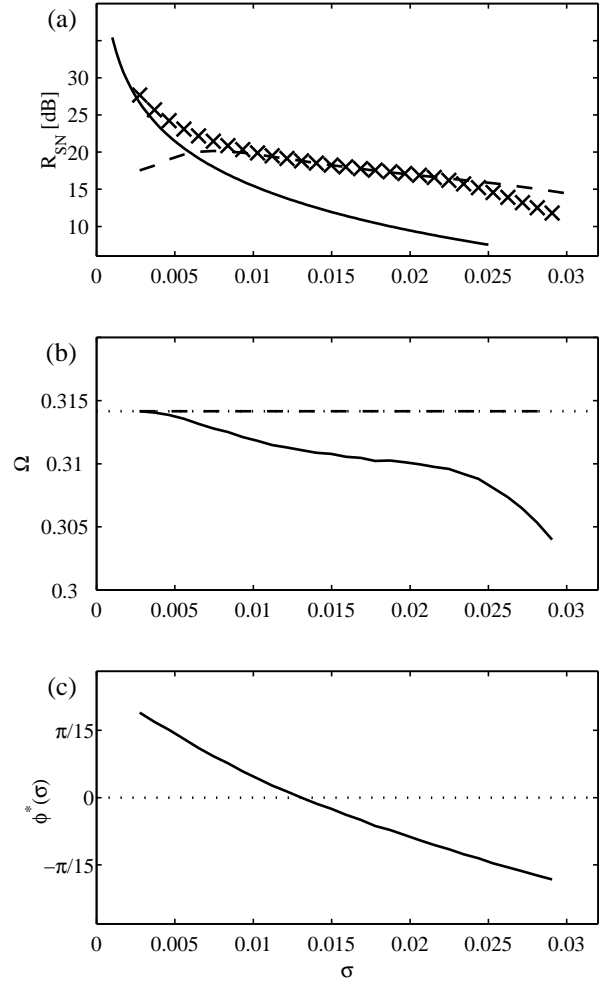}}
  \caption{%
    (a)~Signal-to-noise ratio vs.\ input noise amplitude for adapted
    reset phase $\phi_0=\phi^*(\sigma)$ (symbols) and for fixed reset
    phase $\phi_0=0$ for comparison (dashed).  The solid line is the
    approximation of Eq.~(\ref{eq:snrgauss}).
    (b)~Peak frequency $\Omax$ (solid) and reset frequency $\Omres$
    (dashed) vs.\ noise amplitude for fixed reset phase.  The dotted
    line marks the nominal stimulus frequency.
    (c)~Preferred frequency $\phi^*(\sigma)$ vs.\ noise.
    Stimulus parameters as in Fig.~\ref{fig:stim}.}
  \label{fig:sr_opt}
\end{figure}

A LIF neuron driven by a sinusoidal stimulus which is \emph{not} reset
after each spike will approach a stationary firing pattern
\cite{Tate:1995(917),Ples:1999(7008)}.  The preferred firing phase in
the stationary state is given by Eq.~(\ref{eq:pref}) and depends on
the noise amplitude as shown in Fig.~\ref{fig:sr_opt}(c): the neuron
fires at later phases for weaker noise.  Note that interspike intervals
will be multiples of the stimulus period $T$ in this regime, since the
neuron fires all spikes at the same phase $\phi^*$ (up to jitter).

This observation suggests how to construct a proper approximation to
the full LIF dynamics using the reset model: for each stimulus and
each noise amplitude, determine the preferred phase from
Eq.~(\ref{eq:pref}) and use this phase as reset phase,
\begin{equation}
  \phi_0=\phi_0(\sigma; \mu, q, \Omega) = \phi^*(\sigma; \mu, q,
  \Omega) .
  \label{eq:popt}
\end{equation}
The stimulus is then reset in intervals of multiples of the stimulus
period, so that the effective stimulus will differ from the nominal
stimulus only through jitter.  This jitter vanishes as input noise
vanishes, whence perfect periodic stimulation will be attained for
$\sigma \to 0$.  Consequently, the signal-to-noise ratio will diverge
for vanishing noise as shown in Fig.~\ref{fig:sr_opt}(a).  The reset
frequency is identical to the nominal frequency by construction,
$\Omres=\Omega$, cf.\ Fig.~\ref{fig:sr_opt}(b).  The peak frequency,
on the other hand, converges to the stimulus frequency as noise
vanishes, $\Omax\to\Omres=\Omega$, as the effective stimulus becomes
identical to the nominal one.

We shall now make this argument rigorous.  The ISI distribution of a
LIF model neuron is well approximated by \cite{Ples:2000(367)}
\begin{equation}
  \rho(\tau) \approx h(\tau) \exp\left[-\int_0^{\tau} h(s) ds\right] 
  \label{eq:rhoh}
\end{equation}
\begin{equation}
  h(\tau) =
  \exp\left[-\left(\frac{1-v_0(\tau)}{\sigma}\right)^2\right]
  \label{eq:h}
\end{equation}
where $v_0(\tau)$ is the membrane potential in the absence of noise,
i.e., the solution of Eq.~(\ref{eq:v}) for $\sigma=0$.  We consider
the limit of small noise ($\sigma\ll 1$) and slow stimulation ($T\gg
1$). In that case, we can neglect transient terms in the membrane
potential to obtain
\begin{equation}
  v_0(\tau) = \mu + \hat{q}\cos(\Omega\tau+\Omega\zeta) 
       +{\mathcal O}(e^{-\tau}) ,
  \label{eq:v0}
\end{equation}                 
where $\hat{q}=1/\sqrt{1+\Omega^2}$, and
$\zeta=(\phi_0-\arctan\Omega)/\Omega$.  The hazard $h(\tau)$ will thus
be a sequence of narrow peaks around the maxima of $v_0(\tau)$, i.e.\ 
around $\tau_n = nT-\zeta$.  ISI lengths are multiples of the stimulus
period for adapted reset phase by construction, whence $\zeta=0$ for
$\phi_0=\phi^*(\sigma)$. 

\begin{figure}
  \centerline{\epsfbox{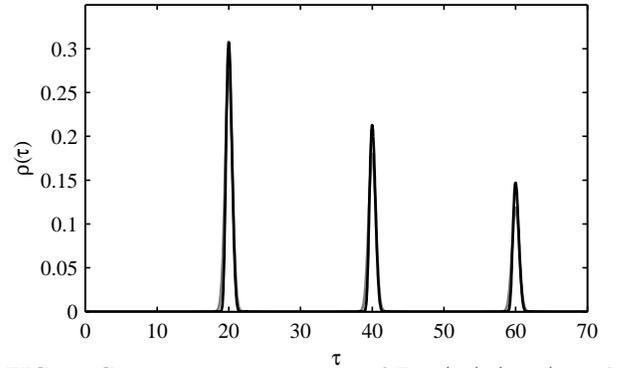}}
  \caption{%
    Gaussian approximation of Eq.~(\ref{eq:rhogauss}) (grey) to the
    interspike-interval distribution (black) for the same stimulus
    parameters as before and $\sigma=0.0046$.}
  \label{fig:gauss}
\end{figure}

As peaks are narrow, we may take the exponential term in
Eq.~(\ref{eq:rhoh}) to be constant across each peak; it merely reduces
the peak amplitude by a factor $\gamma$ for each subsequent peak.  The
ISI distribution can thus be approximated as a superposition of
dampened copies of the hazard function centered about $\tau_n=nT$, i.e.,
\begin{equation}
  \rho(\tau) \approx \sum_{n=1}^{\infty} \gamma^{n-1}
    \exp\left[-\left(\frac{1-v_0(\tau-\tau_n)}{\sigma}\right)^2\right]
  \label{eq:rhosum}
\end{equation}
with $\gamma=\exp\left(-\int_{T/2}^{3T/2} h(s) ds\right)$.  
Expanding the exponent separately for each summand
and retaining only terms of lowest order in $\tau-\tau_n$ yields,
\begin{equation}
  \rho(\tau) \approx c \sum_{n=1}^{\infty} \gamma^{n-1}
       \exp\left(-\frac{\Omega^2(\tau-nT)^2}{\eta\sigma^2}\right) ,
  \label{eq:rhogauss}
\end{equation}
where $\eta^{-1} = \hat{q}(1-\mu-\hat{q})$, and $c$ is a normalization
factor.
The interspike-interval distribution has thus been reduced to a sum of
Gaussians.  This approximation holds well for small noise, as shown in
Fig.~\ref{fig:gauss}.  

The Fourier transform of the ISI distribution is thus
\begin{equation}
  \tilde{\rho}(\omega) = 
    \frac{(1-\gamma)\exp\left[
      -\frac{\eta\sigma^2}{4}\left(\frac{\omega}{\Omega}\right)^2
         + 2\pi i \frac{\omega}{\Omega} \right]}%
         {1-\gamma\exp\left[2\pi i \frac{\omega}{\Omega}\right]} .
  \label{eq:rhofgauss}
\end{equation}
In the limit of small noise and for adapted reset phase, the effective
stimulus becomes equal to the nominal stimulus, and the spectral power
will be maximal at the nominal stimulus frequency in this limit, i.e.,
$\hat{\Omega}\to\Omega$ for $\sigma\to0$, cf.\ 
Fig.~\ref{fig:sr_opt}(b).  The signal-to-noise ratio is therefore 

\begin{equation}
  \rsn =       \frac{S(\Omega)}{S_P} 
       \approx \coth \frac{\eta\sigma^2}8 \quad (\sigma\ll 1)
  \label{eq:snrgauss}
\end{equation}
in good agreement with numerical results for the exact model (solid
line in Fig.~\ref{fig:sr_opt}(a)).  We find in particular that the
signal-to-noise ratio diverges for vanishing noise in the Gaussian
approximation.


\section{Conclusions}\label{sec:disc}

The aim of this paper was to clarify whether the response of the leaky
integrate-and-fire neuron to periodic subthreshold stimulation can be
approximated as a renewal process.  In particular, we wanted to know
whether stochastic resonance at weak noise as reported in earlier work
\cite{Ples:1997(228),Shim:1999(R33)} is a genuine property of the LIF
neuron, or rather an artefact of the stimulus reset, which was
introduced to reduce the full neuronal dynamics to a renewal process.

We argued that renewal (endogenous) dynamics are a good approximation
to the full (exogenous) dynamics only if the phase $\phi_0$, to which
the stimulus is reset after each spike, is adapted to the stimulus
parameters, especially the noise amplitude, in such a way that the
neuron fires most likely at phase $\phi_0$.  We showed that stochastic
resonance does not occur in this case.  Stochastic resonance is only
found if the reset phase is held fixed as the amplitude of the input
noise is varied.

There is no biologically plausible way in which a neuron could reset a
stimulus impinging on it to a fixed phase.  L{\'a}nsk{\'y} suggested
that neurons driven by internally generated membrane potential
oscillations could reset their oscillator to a fixed phase, hence the
term endogenous stimulation \cite{Lans:1997(2040)}.  Cold receptor
cells are driven by internal subthreshold oscillations, but do not
reset their internal oscillator upon firing \cite{Long:1996(215)}.  It
is thus highly unlikely to find neurons which reset the oscillator
that drives them to a fixed phase independent of stimulus properties,
i.e., follow genuine endogenous dynamics.  This in turn means that any
effect arising solely from the reset to a fixed stimulus phase will
not be found in real neurons.  The stochastic resonance effect
reported in Ref.~\cite{Ples:1997(228)} (see also
\cite{Shim:1999(R33)}) is thus a model artefact.

We stress that this conclusion applies only to the particular type of
stochastic resonance discussed here.  The leaky integrate-and-fire
neuron benefits from stochastic resonance when encoding periodic
signals \emph{without} reset (exogenous stimulation), as shown in
\cite{Shim:1999(3461),Ples:1999(7008),Shim:1999(R33)}.  This resonance
occurs at noise amplitudes which are about one order of magnitude
larger than the resonance found in the model with reset to a fixed
phase.  The crucial difference is that studies on exogenous
stimulation consider either explicitly or implicitly the power
spectral density of a spike train of finite duration, while an
infinite spike train is assumed here, cf.\ Sec.~\ref{sec:snr}.  Trains
of very low intensity, but precisely phase-locking to the stimulus, as
found for very weak noise, yield a small finite-time signal-to-noise
ratio, while their infinite-time SNR may be large.

Other noise-induced resonance phenomena found in the LIF with stimulus
reset, e.g., in the interspike-interval distribution or the mean ISI
length \cite{Buls:1996(3958),Buls:1994(4989)}, are not directly
affected by our finding either.  Indeed, Shimokawa et al.\ found
comparable resonance effects in terms of the ISI distribution for
stimulation with and without reset \cite{Shim:1999(3461)}.  These
resonances occur at much larger noise amplitudes than studied here;
see also Ref.~\cite{Shim:1999(R33)}.

Our findings should, on the other hand, not be restricted to the
particular neuron model studied here. We expect that stochastic
resonance may be introduced to nearly any threshold system when the
full dynamics under periodic forcing are reduced to a renewal process
by resetting the forcing to a noise-independent phase after each
threshold crossing.  Periodic forcing may then be recovered by
adjusting the noise such that threshold crossings occur around the
predefined reset phase, whence periodic forcing is recovered, and
optimal output attained.  Unless the fixed reset phase has a
counterpart in the physical system under study, this resonance is
obviously an artefact of a simplification carried too far.


\acknowledgments We would like to thank G.~T.~Einevoll for critically
reading an earlier version of the manuscript and two anonymous
referees for helpful comments. HEP is supported by an EU
Marie-Curie-Fellowship.





%
%

\end{document}